\documentclass[english,aps,manuscript,aps,prb,reprint,showpacs]{revtex4-1}
\usepackage[T1]{fontenc}
\usepackage[latin9]{inputenc}
\usepackage{color}
\usepackage{babel}
\usepackage{amsmath}
\usepackage{graphicx}
\usepackage{esint}
\usepackage[unicode=true,pdfusetitle,
 bookmarks=true,bookmarksnumbered=false,bookmarksopen=false,
 breaklinks=false,pdfborder={0 0 0},backref=false,colorlinks=true]
 {hyperref}
\hypersetup{
 citecolor=blue}

\makeatletter

\@ifundefined{textcolor}{}
{%
 \definecolor{BLACK}{gray}{0}
 \definecolor{WHITE}{gray}{1}
 \definecolor{RED}{rgb}{1,0,0}
 \definecolor{GREEN}{rgb}{0,1,0}
 \definecolor{BLUE}{rgb}{0,0,1}
 \definecolor{CYAN}{cmyk}{1,0,0,0}
 \definecolor{MAGENTA}{cmyk}{0,1,0,0}
 \definecolor{YELLOW}{cmyk}{0,0,1,0}
}

\makeatother

\begin{document}

\title{The resonant dynamics of arbitrarily-shaped meta-atoms}

\author{David A. Powell}

\email{david.a.powell@anu.edu.au}

\affiliation{Nonlinear Physics Centre and Centre for Ultrahigh bandwidth Devices
for Optical Systems, The Australian National University, Canberra
ACT, Australia}
\begin{abstract}
Meta-atoms, nano-antennas, plasmonic particles and other small scatterers
are commonly modeled in terms of their modes. However these modal
solutions are seldom determined explicitly, due to the conceptual
and numerical difficulties in solving eigenvalue problems for open
systems with strong radiative losses. Here these modes are directly
calculated from Maxwell's equations expressed in integral operator
form, by finding the complex frequencies which yield a homogenous
solution. This gives a clear physical interpretation of the modes,
and enables their conduction or polarization current distribution
to be calculated numerically for particles of arbitrary shape. By
combining the modal current distribution with a scalar impedance function,
simple yet accurate models of scatterers are constructed which describe
their response to an arbitrary incident field over a broad bandwidth.
These models generalize both equivalent-dipole and and equivalent-circuit
models to finite sized structures with multiple modes. They are applied
here to explain the frequency-splitting for a pair of coupled split
rings, and the accompanying change in radiative losses. The approach
presented in this paper is made available in an open-source code.
\end{abstract}

\pacs{81.05.Xj, 78.67.Pt, 84.40.Ba}

\maketitle

\section{Introduction}

Resonances are fundamental to many modern photonic and electromagnetic
systems, including metamaterials, nano-antennas, and plasmonic and
dielectric oligomers, all of which seek to strongly manipulate scattering
using small elements. For example, the negative index of a metamaterial
is usually associated with a resonance in the magnetic polarizability
of the constituent meta-atoms. Typical nano-antenna designs consist
of coupled metallic rods operating near their resonance. Fano resonances
arise in plasmonic oligomers due to interference between the modes
of the coupled system. The resonant nature of these systems makes
it highly desirable to create simple oscillator models to describe
their dynamics. Although not necessarily having dimensions much smaller
than the wavelength, the building blocks of these systems are typically
\emph{not large compared to the wavelength}, thus they can be adequately
described by a small number of modes. 

For metamaterials consisting of a large, three-dimensional array,
much effort has been dedicated to homogenization approaches, whereby
the metamaterial is approximated by a continuous medium, and the system
is described in terms of average fields \cite{Simovski2011}. However,
in many systems of interest, the required criteria for homogenizability
are not satisfied, either because the meta-atoms are not sufficiently
sub-wavelength, the arrays are so small that boundary effects and
radiation losses are very strong, or the arrangement is not periodic.
As an alternative to homogenization, it is possible to consider the
fields of a metamaterial's Bloch modes as the fundamental degrees
of freedom \cite{Zeng2012a}, however this suffers from many of the
same limitations. In many cases the modes of \emph{individual resonators}
form a much more convenient basis to study the behavior of meta-atoms
and resonant scatterers, since the number of excited modes is typically
small. For many experimental configurations reported in the literature,
the number of scatterers in the system is small enough that it is
feasible to explicitly describe the modal excitation of each of them. 

\begin{figure}[t]
\includegraphics[width=1\columnwidth]{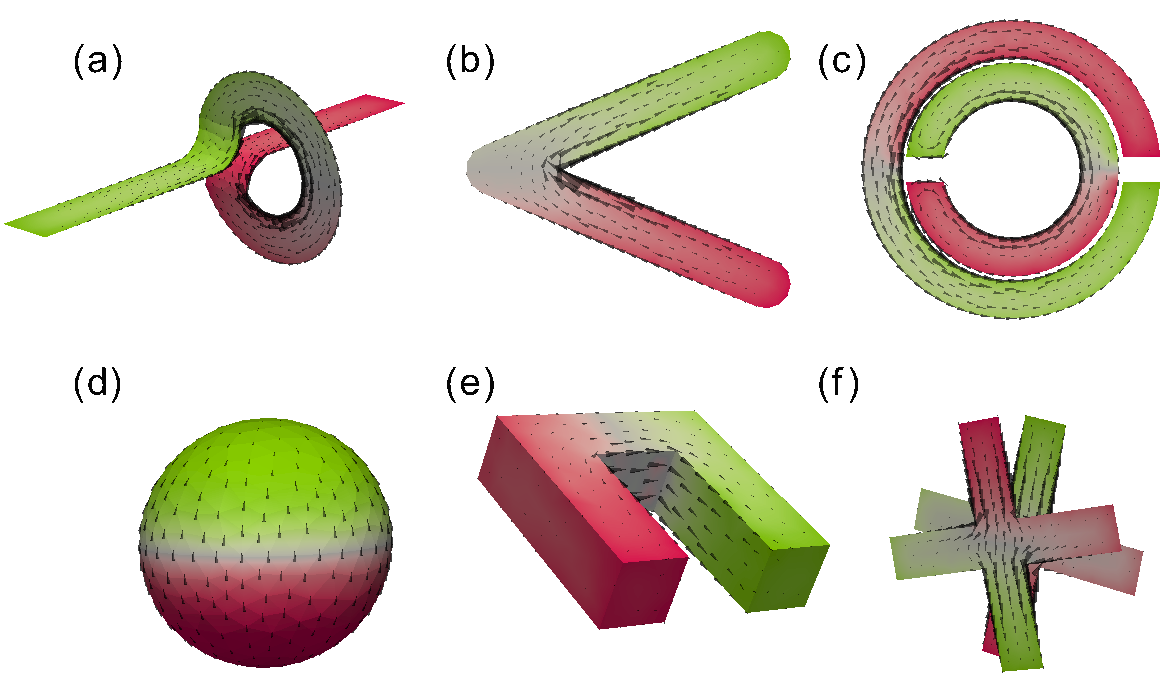}

\protect\caption{(Color online) Meta-atoms each showing charge (colors) and current
(arrows) distributions for one of their modes. (a) Canonical spiral,
(b) V-antenna, (c) split-ring resonator, (d) sphere, (e) horseshoe
and (f) twisted crosses.\label{fig:montage}}
\end{figure}

\emph{In this article, the modes of arbitrarily-shaped resonant particles
are found and are used to construct simple oscillator models.} These
models give a highly accurate description of the particles over a
very broad bandwidth and directly include radiative effects. They
satisfy causality, and can account for the coupling between particles,
which gives rise to hybridized modes. Examples of applicable structures
and their modes are shown in Fig.~\ref{fig:montage}. For all of
these structures the fundamental modes are plotted, except in Fig.~\ref{fig:montage}(f),
where a second-order mode is shown. Routines to calculate the coefficients
of the oscillator model from the scatterer geometry are implemented
in an open-source software package \texttt{OpenModes} \cite{openmodes}. 

This paper is organized as follows. In Section \ref{sec:modes-open-resonators},
existing approaches to find the modes of open resonators are discussed,
leading to the proposed approach which is presented in Section \ref{sec:model-single}.
In Section \ref{sec:coupling-meta-atoms}, this model is used to solve
the problem of coupled resonators, showing how the influence of multiple
modes of each uncoupled resonator can easily be taken into account
in the hybridization process. Appendix \ref{sec:EFIE-details} gives
details of the EFIE operator used, Appendix \ref{sec:implementation-details}
gives general details of the numerical implementation, and Appendix
\ref{sec:search-procedure} outlines the procedure to find the singularities.

\section{The physics of open resonators and their modes\label{sec:modes-open-resonators}}

The simplest approach to finding the modes of resonant particles is
to illuminate the structure at a frequency corresponding to a resonance,
and observe the fields either through numerical simulation, or with
experimental techniques such as near-field scanning microscopy \cite{Alonso-Gonzalez2011}.
These approaches work well if the modes are greatly separated in frequency,
however many systems operate in a regime of overlap or interference
between different modes \cite{Miroshnichenko2010}, and it is difficult
to distinguish the contribution of each mode to the total response.
To develop a model for the resonances in meta-atoms or plasmonic structures,
a simple and appealing approach is to develop equivalent circuit models
\cite{Bilotti2007,Ikonen2007a,Karyamapudi2005,Staffaroni2012}. This
has the advantage of providing insight and computational simplicity,
and builds on well established results in antenna and microwave theory.
However, such a circuit model must be developed manually for each
different meta-atom type, and important physical phenomena such as
coupling to radiated fields are not adequately described by lumped
circuit elements.

Models which represent the particles by their dipole moments have
been constructed for meta-atoms, and these can gives a reasonable
description of the far-field coupling of the fundamental modes, including
radiation effects \cite{Sersic2011}. The most well-studied case using
dipole or multipole methods is that of a sphere embedded in a uniform
background, since it has a multipole solution which describes the
polarization of each mode in closed form \cite{Bohren1983}. However,
many practical systems have resonators which are of much lower symmetry,
either through deliberate design or the presence of a perturbing substrate,
and the modes of complex-shaped resonators can have significant contributions
from many multipole terms \cite{Krasnok2013a}. Particularly for near-field
interaction effects, all of these higher-order terms must be included
in calculations, thus the multipole approach loses its simplicity.
To model such systems effectively, analytical methods cannot be used
to find the eigenmodes, and some other solution must be found. For
plasmonic particles, it is possible to find the modes numerically
under the quasi-static approximation \cite{Mayergoyz2005}. This approach
makes very strong assumptions about the sub-wavelength nature of the
particle, and radiation effects are neglected, although they may be
added back to the model as a perturbation \cite{Davis2009}.

In closed cavities, modes can be found by expressing Maxwell's equations
in eigenvalue form, with the eigenvalues corresponding to the resonant
frequencies. However, meta-atoms and nano-antennas are intrinsically
open systems, which radiate into the surrounding environment. Their
modal near fields are not strictly confined to any well-defined region
of space, and must somehow be disentangled from the radiating fields.
Particularly for plasmonic particles, their may also be strong dissipative
losses. Thus it is not appropriate to solve the lossless problem and
to treat radiative and dissipative losses as a perturbation, since
they can induce strong qualitative changes in the response of a system
\cite{Yakovlev2000,Davoyan2010}.

In the language of Hamiltonian mechanics, these significant losses
mean that the system must be described by a non-Hermitian operator.
In fields such as quantum optics, open systems have been studied using
the ``system and bath'' approach \cite{Viviescas2003}. In this
model, the system is partitioned into a resonant system and a continuum
of modes, and coupling terms between the two are introduced. Although
this description is complete, one drawback is that the partitioning
of the system is not unique, thus the modes are not uniquely defined.
Additionally, it is necessary to include an infinite continuum of
plane waves into the calculations, which is somewhat cumbersome, and
cannot be considered as a simple oscillator model. Similar models
incorporating the complete spectrum of plane waves have been utilized
for metamaterials \cite{Jenkins2012}, under the assumption that each
meta-atom is described by a single electric and magnetic dipole moment.

An alternative approach is to study the quasi-normal modes, which
are self-consistent undriven solutions occurring at complex values
of frequency. They extend the familiar concept of resonant modes to
dissipative systems, and although such modes are not orthogonal in
the usual sense, in certain cases they do satisfy orthogonality over
an unconjugated inner product \cite{Leung1994}. Such modes seem to
offer an intuitive description for resonant systems, however, not
only are they not-well confined, their fields actually \emph{diverge}
with increasing distance from the resonator. This corresponds to temporal
solutions as $t\rightarrow\infty$, where almost all energy has escaped
from the cavity into radiated fields \cite{Kristensen2012}. By utilizing
appropriate absorbing boundary conditions this divergence can be handled
\cite{Sauvan2013,bai_efficient_2013}, however this spatially divergent
field is an inconvenient representation of a compact object. In Ref.~\onlinecite{bai_efficient_2013}
it was shown that absorption, scattering and emission effects can
be calculated from quasi-normal modes, and it is noteworthy that the
all the relevant formulas effectively integrate the polarization current
over the volume of the scatterer.

Integral equation approaches which solve for currents are routinely
used in scattering theory, and the singularities of the scattering
operator can also yield solutions\cite{Hanson2002}, which are essentially
the same as quasi-normal modes. In Ref.~\onlinecite{Kovalyov2011}
these were calculated based on a spherical harmonic decomposition.
This approach is well suited to modeling antennas or scatterers separated
by relatively large distances, and all far-field radiation channels
are explicitly incorporated. However, spherical harmonics are is not
suitable for modeling the strongly varying near-fields which couple
meta-atoms together, nor the influence of an inhomogeneous background,
and this also involves an arbitrary and non-physical partitioning
of space into internal and external parts. For open resonators which
are uniform along one direction, a comprehensive theory was presented
in Ref.~\onlinecite{Shestopalov1996}, however this restriction excludes
many structures of practical interest.

The advantage of the scattering approach is that the solution is given
in terms of the conduction or polarization current only within the
resonant structure. This gives a more useful description of the mode,
since currents remains finite at complex frequencies, in contrast
to the corresponding fields which diverge. The natural approach to
solving for the current on the resonator directly is to use integral
equation approaches, known variously in the literature as the method
of moments \cite{Harrington1968,Gibson2008}, boundary element method
or integral equation method. In these approaches, the current is expanded
into a finite number of basis functions, and a Green's function is
used to calculate the interaction between all the current elements.
By solving the resulting impedance matrix, the solution can be found
for any external exciting field, and the radiation boundary conditions
are automatically taken into account. Such approaches are well-established
for solving microwave scattering problems, and more recently they
have been applied successfully to dielectric and plasmonic nano-structures
\cite{Kern2009,Ergul2012,Zheng2013a}. It is important to emphasize
that while these approaches use the terminology of impedance taken
from circuit theory, it was shown in Ref.~\onlinecite{greffet_impedance_2010}
that it has an alternative interpretation related to the local density
of states. 

Modes of the structure correspond to the singularities of the impedance
matrix in the complex plane, and these have been found for dielectric
resonators \cite{glisson_evaluation_1983}, and have been used to
describe the transient scattering of a radar pulse from a target \cite{Baum1976,Baum1978}.
This approach can be understood as applying analytical continuation
to the eigenvalue expansion, which provides a scalar description of
the structure, but which must be recalculated at each frequency of
interest. In Ref.~\onlinecite{Zheng2013a} such an eigenvalue expansion
was applied the impedance matrix in order to calculate the excitation
and coupling of plasmonic dolmen structures. In Ref.~\onlinecite{makitalo_modes_2014}
the properties of the Müller formulation of the surface integral problem
were studied in detail, and it was shown that the singularities of
the integral operator can yield the resonances of plasmonic structures
in both full-wave and quasi-static regimes. In Ref.~\onlinecite{zheng_line_2013}
the location of the singularities of plasmonic structures in the complex
frequency was related to the quality factor and the stored energy
in the near-fields of the structure. In the next section a procedure
for finding these singularities will be presented, and they will be
used to develop a simple oscillator model accounting for all excitation,
coupling, radiation and interference effects.

\section{Modeling a single element\label{sec:model-single}}

In this section a numerical model is constructed for an individual
resonant element. It is then shown how analyzing the frequencies where
the impedance matrix is singular yields a compact model which is accurate
over a broad bandwidth, and describes each mode with quite simple
dynamics.

\subsection{The electric field integral equation}

All dynamic quantities have implicit time dependence of $\exp(st)$
with $s=\Omega+j\omega$, and are related to time domain quantities
via a two-sided Laplace transform pair \cite{Baum1978}. The electric
field $\mathbf{E}_{s}$ scattered by an object is related to its induced
currents $\mathbf{j}$ via the electric field integral equation (EFIE):

\begin{equation}
\mathbf{E}_{s}\left(\mathbf{r},s\right)=\iiint_{\Gamma}\overline{\overline{G}}_{0}(\mathbf{r}-\mathbf{r}',s)\cdot\mathbf{j}(\mathbf{r}',s)\mathrm{d^{3}\mathbf{r}}\label{eq:EFIE}
\end{equation}
where the free space dyadic Green's function $\overline{\overline{G}}_{0}$
is given by Eq.~\eqref{eq:greens_function}, and $\Gamma$ is the
volume of the object. For perfectly conducting metals considered here,
the the tangential components of the scattered field and incident
field $\mathbf{E}_{i}$ cancel on the surface $\hat{\mathbf{n}}\times\mathbf{E}_{i}=-\hat{\mathbf{n}}\times\mathbf{E}_{s}$,
the integration is over the object surface $\partial\Gamma$ and the
resulting operator equation is denoted $\mathbf{E}_{i}=\mathcal{Z}\left(\mathbf{j}\right)$.
More general formulations can include polarization within dielectrics
or imperfect metals, through a volume \cite{Zheng2013a} or surface
equivalent problem \cite{Yla-Oijala2005}. Furthermore, if a different
Green's function is used in Eq.~\eqref{eq:EFIE}, background media
can be incorporated while still only solving for currents on the scatterer,
with layered media \cite{Michalski1997} being of particular interest.
Note that artificial magnetism due to circulating currents is accounted
for in Eq.~\eqref{eq:EFIE} by the gradient of the electric field
\cite{simovski_effective_2010}, without requiring the additional
terms used in some models \cite{Jenkins2012,Sersic2011}. 

In Ref.~\onlinecite{Hanson2002} the properties of such integral
operators are discussed in detail, using the tools of functional analysis.
In particular the spectral properties of such operators are discussed,
which are relevant to the techniques used in this section. Although
many of the relevant proofs are not directly applicable to Eq.~\eqref{eq:EFIE},
the uniqueness theorem means that the solutions found are genuine
physical properties which are independent of the particular formulation
of Maxwell's equations which is used. To solve the operator equation
numerically, the geometry is represented by a triangular surface mesh
and the current is expanded into basis functions $\mathbf{f}_{n}$,
each defined over an area $T_{n}$ 

\begin{equation}
\mathrm{\mathbf{j}}\left(\mathbf{r}\right)=\sum_{n=1}^{N}I_{n}\mathbf{f}_{n}\left(\mathbf{r}\right).\label{eq:current_basis}
\end{equation}
To obtain a finite number of equations, the incident field is weighted
by the same set of basis functions 
\[
V_{n}=\iint_{T_{n}}\mathbf{f}_{n}\left(\mathbf{r}\right)\cdot\mathbf{E}_{i}\left(\mathbf{r}\right)\mathrm{d^{2}}\mathbf{r}.
\]
 Applying the same procedures to Eq.~\eqref{eq:EFIE} results in
a matrix equation which describes the full dynamics

\begin{equation}
\mathrm{V}(s)=\mathrm{Z}(s)\cdot\mathrm{I}(s),\label{eq:impedance}
\end{equation}
where $\mathrm{V}$ and $\mathrm{I}$ are vectors of length $N$,
whilst $\mathrm{Z}$ is the $N\times N$ impedance matrix which approximates
the operator $\mathcal{Z}$, and is defined in Eq.~\eqref{eq:impedance_matrix}.

The impedance matrix is closely related to the interaction matrix
in coupled-dipole models \cite{Sersic2011}, but it has the advantage
of directly including both mutual and self-interaction effects. It
is sometimes useful to separate it into two parts according to the
dominant frequency dependence, which is equivalent to separating the
contributions of the scalar and vector potentials in the Lorenz gauge
\begin{equation}
\mathrm{Z}\left(s\right)=s\mathrm{L}\left(s\right)+\frac{1}{s}\mathrm{S}\left(s\right).\label{eq:impedance_LS}
\end{equation}
In the limit $\gamma|\mathbf{r}|\rightarrow0$ (where $\gamma=s\sqrt{\varepsilon\mu}$
is the complex propagation factor), these matrices correspond to the
inductance and elastance (the inverse of capacitance) respectively,
hence the symbols for the corresponding scalar quantities are used
\cite{Kennelly1936}. Due to Lorentz reciprocity and the use of identical
basis and weighting functions, $\mathrm{Z}$ is a complex-symmetric
matrix, in the sense $Z_{mn}=Z_{nm}$, but in general $Z_{mn}\neq\overline{Z}_{nm}$
(the bar denotes complex conjugation).

\subsection{Modes as frequency-dependent eigenvectors\label{sec:EEM}}

From the impedance matrix $\mathrm{Z}(s)$, a simple model can be
extracted at each frequency, by solving the eigenvalue problem

\begin{equation}
\mathrm{Z}(s)\cdot\mathrm{I}^{(\alpha)}\left(s\right)=z^{(\alpha)}\left(s\right)\mathrm{G}\cdot\mathrm{I}^{(\alpha)}\left(s\right),\label{eq:impedance_eigenproblem}
\end{equation}
where the eigenvalue $z^{(\alpha)}(s)$ is a scalar impedance, and
the eigenvector $\mathrm{I}^{(\alpha)}\left(s\right)$ gives the corresponding
current distribution. The matrix equation~\eqref{eq:impedance_eigenproblem}
is a numerical approximation of the operator eigenvalue equation $\mathcal{Z}\left(\mathbf{j}^{\left(\alpha\right)}\right)=z^{(\alpha)}\mathcal{I}\left(\mathbf{j}^{\left(\alpha\right)}\right)$,
where $\mathcal{I}$ is the identity operator. Since the basis and
testing functions are not orthonormal, the matrix form of the identity
operator is the Gram matrix\cite{Harrington1968,Bekers2009} $\mathrm{G}$,
where $G_{mn}=\int_{T_{m}}\int_{T_{n}}\mathbf{f}_{m}\left(\mathbf{r}\right)\cdot\mathbf{f}_{n}\left(\mathbf{r}\right)\mathrm{d^{2}}\mathbf{r}$.
Some works \cite{Zheng2013a,Baum1978} use the identity matrix for
this term, thus the eigenvalues may have have contributions related
to the mesh density, in addition to the those from the dynamics of
the physical system.

\begin{figure}
\includegraphics[width=1\columnwidth]{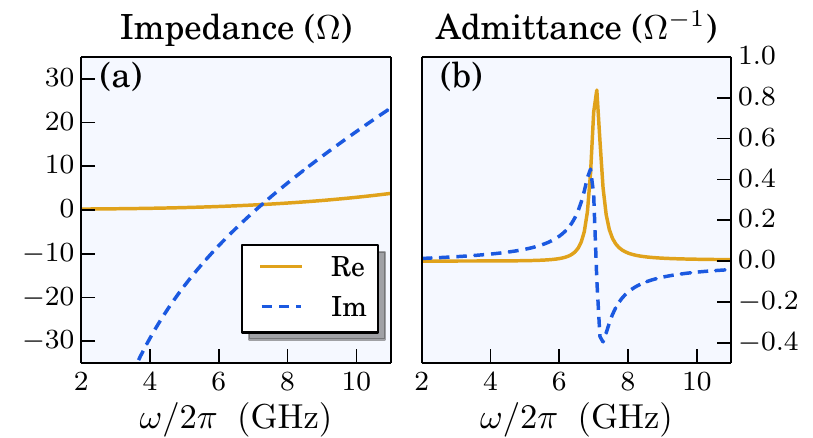}

\protect\caption{(Color online) (a) Eigenvalue for the first mode of an SRR, having
dimensions of impedance, and (b) its inverse which has dimensions
of admittance, with a Lorentzian-like shape.\label{fig:eigenvalue-imaginary-axis}}
\end{figure}

Once the eigenvectors are scaled such that they satisfy the orthonormality
relationship

\begin{equation}
\iint\limits _{\partial\Gamma}\mathbf{j}^{(\alpha)}\left(\mathbf{r}\right)\cdot\mathbf{j}^{(\beta)}\left(\mathbf{r}\right)\mathrm{d^{2}}\mathbf{r}=\mathrm{I^{(\alpha)}}^{\mathrm{T}}\left(s\right)\cdot\mathrm{G}\cdot\mathrm{I}^{(\beta)}\left(s\right)=\delta_{\alpha\beta},\label{eq:orthonormality}
\end{equation}
the impedance matrix has the decomposition \cite{Hanson2002}

\begin{equation}
\mathrm{Z}(s)=\sum_{\alpha=1}^{N}z^{(\alpha)}\left(s\right)\mathrm{I^{(\alpha)}}\left(s\right)\otimes\mathrm{I}^{(\alpha)}\left(s\right),\label{eq:eigendecomposition}
\end{equation}
where in practice only $\tilde{N}$ modes ($1\leq\tilde{N}\ll N$),
contribute significantly to the response, and in many cases $\tilde{N}=1$
is sufficient. An arbitrary current is decomposed into a series of
modes by the projection dyad $\mathrm{I^{(\alpha)}}(s)\otimes\mathrm{I}^{(\alpha)}(s)$
(without complex conjugation), and the mode dynamics are given by
the impedance $z^{(\alpha)}(s)$. This could be understood as the
sum of equivalent circuit responses, however it is important to emphasize
the role of the projection dyad, which has no equivalent circuit counterpart.
Due to the non-Hermitian nature of the system, the projection of the
incident field onto this dyad can also contribute phase terms. Although
counter-intuitive, this effect is completely physical and accounts
for interference between modes\cite{Hopkins2013}.

To give a detailed example of this model, a single ring split-ring
resonator (SRR) is considered, with inner radius 2.5\ mm, outer radius
4\ mm, gap width 1\ mm, modeled as a thin PEC layer divided into
852 triangles. The dynamics of its first mode are governed by the
eigenvalue $z^{(1)}(s)$ plotted in Fig.~\ref{fig:eigenvalue-imaginary-axis}(a).
It can be seen that the imaginary part of the impedance dominates,
corresponding to the reactive stored energy, while the real part is
much smaller, indicating that the radiation losses for this mode are
relatively low. This is confirmed by Fig.~\ref{fig:eigenvalue-imaginary-axis}(b),
which shows the corresponding admittance $1/z^{(1)}(s)$, more clearly
indicating the resonant nature of the mode and its relatively high
quality factor.

The main drawback of this eigenvalue expansion is that it requires
the full impedance matrix to be calculated and an eigenvalue decomposition
to be performed at every frequency, thus it has the same computational
requirements as a fully numerical model.

\subsection{Modes as singularities of the operator equation}

\begin{figure}
\includegraphics[width=1\columnwidth]{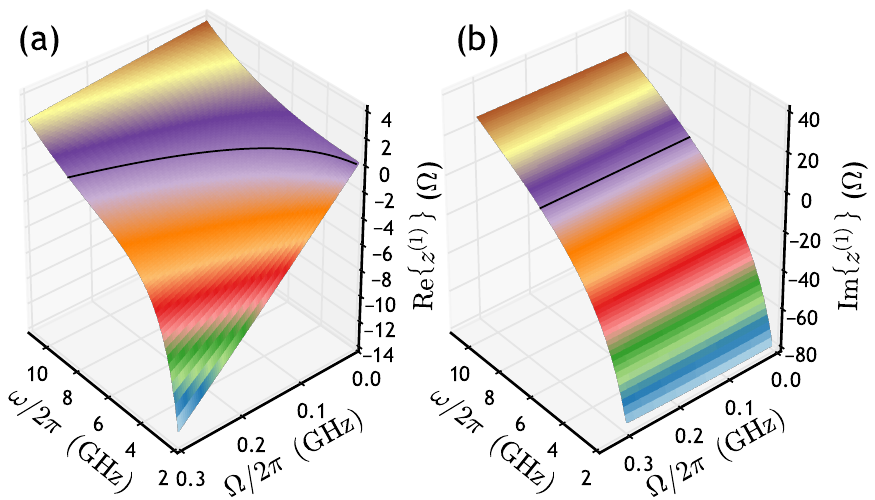}

\protect\caption{(Color online) (a) Real and (b) imaginary parts of the first eigenvalue
of an SRR, plotted as a function of complex frequency. The solid lines
show where each part is zero. Their intersection gives the complex
eigenfrequency.\label{fig:eigenvalues-complex-plane}}
\end{figure}

To develop a single model for a finite scatterer which is accurate
over a wide frequency range, $z^{(\alpha)}(s)$ is extended analytically
into the complex $s$ plane. Fig.~\ref{fig:eigenvalues-complex-plane}
extends the eigenvalue in Fig.~\ref{fig:eigenvalue-imaginary-axis}(a)
in this manner. Its real and imaginary parts are given by the heights
of the surfaces, with the black line indicating where they pass through
zero. Clearly the eigenvalue goes to zero at the intersection of these
two curves. At such frequencies, the impedance matrix $\mathrm{Z}$
is singular \cite{Baum1978}, corresponding to a current solution
which can be sustained without any driving field {[}i.e.~$\mathrm{V}=0$
in Eq.~\eqref{eq:impedance}{]}. In Ref.~\onlinecite{Marin1973}
it was proven that for sufficiently smooth objects all such singularities
are poles, and it was observed that in practice they are of first
order, and no other terms are required to describe the dynamics.

\begin{figure}
\includegraphics[width=1\columnwidth]{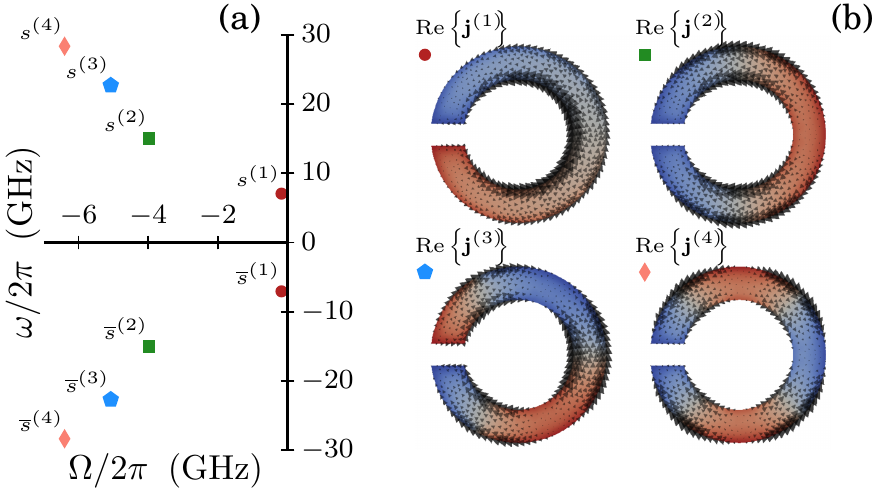}

\protect\caption{(Color online) (a) The complex singularities of a single split-ring
resonator which give the resonant frequencies, and (b) the corresponding
charge and current distributions.\label{fig:singularities}}
\end{figure}

Since the impedance matrix $\mathrm{Z}$ has explicit dependence on
$s$, the root-finding problem is nonlinear in $s$ and must be solved
using iterative methods. A procedure was developed based on robust
starting estimates, as detailed in Appendix \ref{sec:search-procedure}.
For the example presented here, the solution converged to a relative
accuracy of $10^{-8}$ within 10 iterations, making the process quite
efficient. The solution of this nonlinear problem was empirically
found to be more reliable than the solving frequency-dependent linear
eigenvalue problem in Eq.~\eqref{eq:impedance_eigenproblem}, in
the sense that it is much less prone to converge on a spurious non-physical
eigenvalue. In Fig.~\ref{fig:singularities}(a) the singular points
are shown for the first four modes of a split ring resonator. It can
be seen that the higher order modes have have larger real parts of
$s^{\left(\alpha\right)}$, indicating that they have stronger radiative
losses. These singularities occur in complex-conjugate pairs, as the
system response is real in the time domain.

For each frequency $s^{(\alpha)}$ corresponding to the resonance
of a mode, there is a vector $\mathrm{I^{(\alpha)}\left(s^{(\alpha)}\right)}$
which satisfies the homogeneous equation. From this vector, the mode's
current distribution $\mathbf{j}^{(\alpha)}\left(s^{(\alpha)},\mathbf{r}\right)$
is calculated using Eq.~\eqref{eq:current_basis}, and the charge
distribution is given by $q^{(\alpha)}\left(s^{(\alpha)},\mathbf{r}\right)=-\frac{1}{s^{(\alpha)}}\nabla\cdot\mathbf{j}^{(\alpha)}\left(s^{(\alpha)},\mathbf{r}\right)$.
The real parts of these surface charges (colors) and currents (arrows)
are shown in Fig.~\ref{fig:montage} for several different structures.
Despite the strong differences in geometry, it is clear that the lowest-order
modes of canonical spiral, V-antenna, sphere and horseshoe all exhibit
electric dipole-like charge distributions. In Fig.~\ref{fig:singularities}(b)
the first four modes of the single split ring resonator shown, corresponding
to the singularities shown in Fig.~\ref{fig:singularities}(a). It
can be seen that higher order modes exhibit increasing degree of spatial
oscillation with increasing order, similar to the case for simple
closed cavities.

These current distributions are closely linked to the frequency-dependent
eigenvectors discussed in Section \ref{sec:EEM}, which can be regarded
as their analytical continuation away from the singular points in
the $s$ plane. In Ref.~\onlinecite{Andrew1993} it was shown that
careful normalization of the eigenvectors is required for them to
be analytic, and the normalization given in Eq.~\eqref{eq:orthonormality}
satisfies this requirement. For the well-studied case of a sphere,
the eigenvectors are frequency independent \cite{Baum1976}, although
this is not true in the general case

\subsection{The broadband model}

The value in finding singularities in the complex $s$ plane is that
they form a useful basis for modeling the dynamics of the structures.
By numerically evaluating $\mathrm{d}z^{(\alpha)}/\mathrm{d}s$ at
$s=s^{(\alpha)}$, and enforcing the impedance to be open-circuit
at $s=0$, a fourth order model is fitted to each scalar impedance
$z^{(\alpha)}$

\begin{equation}
z^{(\alpha)}\left(s\right)=\frac{z_{-1}^{(\alpha)}}{s}+z_{0}^{(\alpha)}+z_{1}^{(\alpha)}s+z_{2}^{(\alpha)}s^{2}.\label{eq:scalar-model}
\end{equation}

These terms are all real and can be interpreted as elastance (inverse
capacitance), ohmic dissipation, inductance and radiative losses respectively,
and correspond directly to the form of the inverse polarizability
used in dipole models \cite{Sersic2011}. The advantage of this form
is that the correct signs of all terms can be enforced to guarantee
a passive, causal response. Many existing formulations express the
admittance in terms of residues of the poles\cite{Baum1976,zheng_line_2013,Sauvan2013}
of $\mathrm{Z}^{-1}$. Such models can be used instead of Eq.~\eqref{eq:scalar-model},
but will be physically correct only if the residue of the conjugate
pole at $s=\bar{s}^{(\alpha)}$ and the zero at the origin are correctly
accounted for.

\begin{figure}
\includegraphics[width=1\columnwidth]{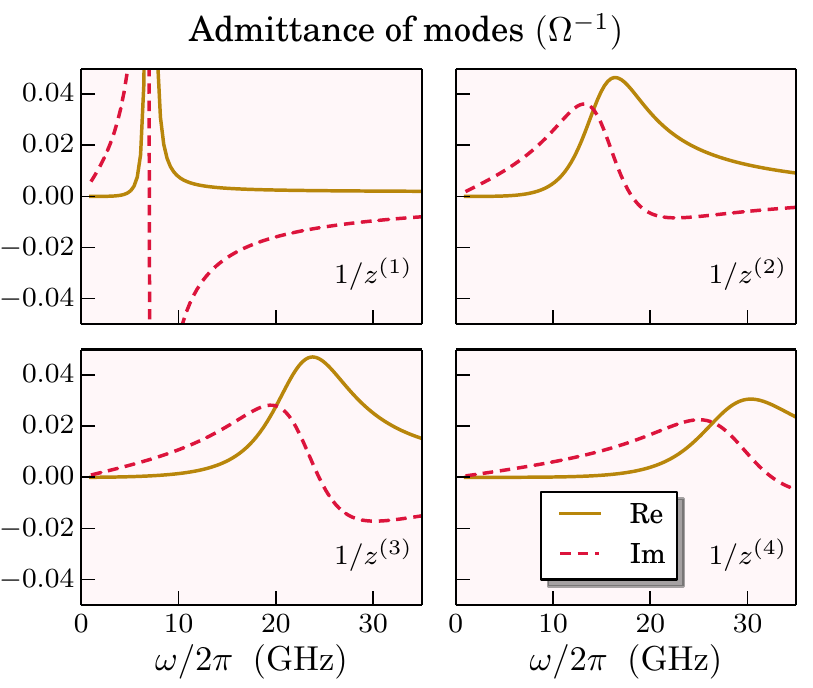}

\protect\caption{(Color online) Scalar admittance functions which describe the dynamics
of the first four modes of a split ring resonator. \label{fig:scalar-admittances}}

\end{figure}

In Fig.~\ref{fig:scalar-admittances} the scalar admittances $1/z^{(\alpha)}(s)$
are plotted corresponding to the modes in Fig.~\ref{fig:singularities}.
The resonant behavior is clearly observable, and is similar to a series
resonant circuit, with the real part reaching the maximum and the
imaginary part crossing through zero at resonance. It can be seen
that the widths of the resonant peaks increase for the higher order
modes, consistent with the increased values of $\Omega$ at the singularities
shown in Fig.~\ref{fig:singularities}(a). It is also clear that
the line-shapes can be highly asymmetric, confirming that the expression
in Eq.~\eqref{eq:scalar-model} is more appropriate than simpler
RLC circuit or Lorentzian models.

While the frequency-dependence of the eigenimpedances $z^{(\alpha)}$
is well characterized by Eq.~\ref{eq:scalar-model}, the analytic
continuation of the vectors $\mathrm{I^{(\alpha)}}\left(s^{\left(\alpha\right)}\right)$
is more subtle. Each $\mathrm{I^{(\alpha)}}(s)$ is an analytic function
of $s$, such that the same eigenvector can be tracked as the frequency
is varied. Since they represent the Laplace transform of a real function,
the currents must obey the the conjugate symmetry relationship $\mathrm{I^{(\alpha)}}(\bar{s})=\mathrm{\overline{I}^{(\alpha)}}(s)$
\cite{Baum1975}. The normalization introduced in Eq.~\eqref{eq:orthonormality}
eliminates the arbitrary complex scaling factor on the eigenvectors,
making it meaningful to distinguish between their real and imaginary
parts. The conjugate symmetry of the eigenvectors requires that the
mode currents are real on the real $s$ axis, including at zero frequency.
Therefore, the presence of a non-zero imaginary part of $\mathrm{I^{(\alpha)}}$
implies that the modal current distributions \emph{cannot be frequency
independent}. However, from a practical point of view, all the current
distributions shown in Figs.~\ref{fig:montage} and \ref{fig:singularities}(a)
have imaginary components (not shown) which are much smaller than
the real parts, allowing their frequency variation to be neglected
with little loss of accuracy.

Utilizing the vector $\mathrm{I^{(\alpha)}}\left(s^{\left(\alpha\right)}\right)$
and corresponding impedance function $z^{(\alpha)}\left(s\right)$,
the response of the particle to an excitation field vector $\mathrm{V}(s)$
is well approximated by: 

\begin{eqnarray}
\mathrm{I}(s) & = & \mathrm{Z}^{-1}(s)\cdot\mathrm{V}(s)\nonumber \\
 & = & \sum_{\alpha=1}^{\tilde{N}}\frac{1}{z^{(\alpha)}\left(s\right)}\mathrm{I^{(\alpha)}}\left(s^{\left(\alpha\right)}\right)\left[\mathrm{I^{(\alpha)}}\left(s^{\left(\alpha\right)}\right)\cdot\mathrm{V}(s)\right],\label{eq:modal-inverse}
\end{eqnarray}

\subsection{Verification}

\begin{figure}
\includegraphics[width=1\columnwidth]{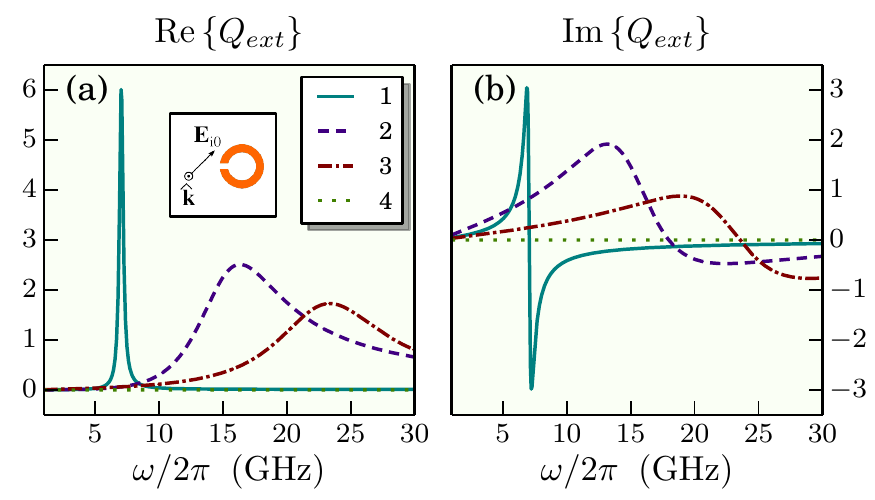}

\protect\caption{(Color online) The contribution of each mode to the (a) real and (b)
imaginary parts of the complex extinction efficiency. The inset gives
the polarization of the incident wave.\label{fig:extinction-modes}}
\end{figure}
 
\begin{figure}
\includegraphics[width=1\columnwidth]{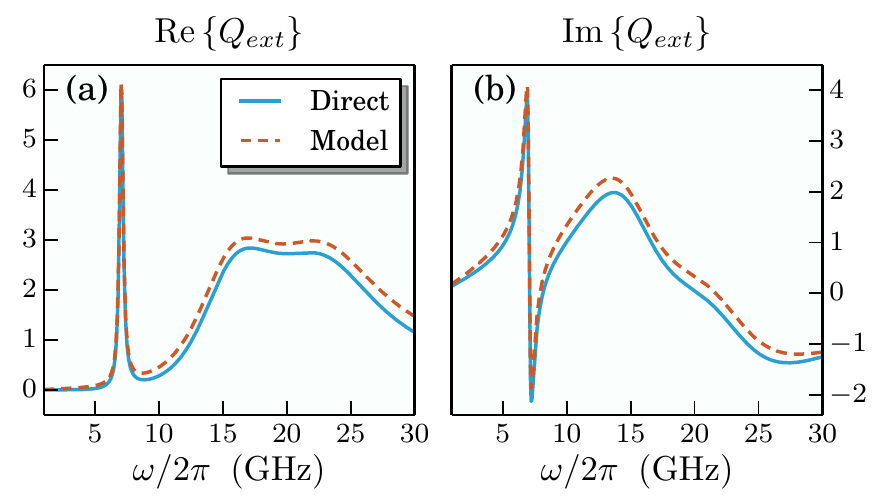}

\protect\caption{(Color online) Total extinction efficiency calculated from the model
and compared with the direct calculation, showing the (a) real and
(b) imaginary parts.\label{fig:extinction-comparison}}
\end{figure}

To verify the accuracy of the model, the solution obtained from Eq.~\eqref{eq:modal-inverse}
is compared with the exact solution of Eq.~\eqref{eq:impedance}.
The simplest quantity which characterizes the response is the extinction
efficiency, which shows the energy extracted from the incident field
by the scatterer. It is calculated as

\begin{eqnarray}
Q_{ext} & = & \frac{\eta\iint_{\partial\Gamma}\mathbf{\overline{E}}_{i}\left(\mathbf{r}\right)\cdot\mathbf{j}\left(\mathbf{r}\right)\mathrm{d^{2}}\mathbf{r}}{\pi r_{o}^{2}\iint_{\partial\Gamma}\mathbf{\overline{E}}_{i}\left(\mathbf{r}\right)\cdot\mathbf{E}_{i}\left(\mathbf{r}\right)\mathrm{d^{2}}\mathbf{r}}\nonumber \\
 & = & \frac{\eta\overline{\mathrm{V}}\left(s\right)\cdot\mathrm{I}\left(s\right)}{\pi r_{o}^{2}\overline{\mathrm{V}}\left(s\right)\cdot\mathrm{V}\left(s\right)},\label{eq:extinction-current}
\end{eqnarray}
where $\eta=\sqrt{\frac{\mu}{\epsilon}}$ is the intrinsic impedance
of the background medium, and $r_{o}$ is the radius of the smallest
sphere enclosing the object %
\footnote{The enclosing sphere is used rather than the more common geometric
cross-section of the object, since the geometric cross-section can
be very small for certain cases such as thin-wire metamaterials%
}. The incident field is a plane-wave described by $\mathbf{E}_{i}\left(\mathbf{r}\right)=\mathbf{E}_{i0}\exp\left(-\gamma\hat{\mathbf{k}}\cdot\mathbf{r}\right)$,
where $\hat{\mathbf{k}}$ is the direction of propagation. In contrast
to the usual definition \cite{bohren_absorption_1983}, both the real
and imaginary parts of extinction are retained. This is done by analogy
with circuit theory, where the complex power delivered to a load is
considered, with the real part corresponding to the time-averaged
power flow, and the imaginary part corresponding to the reactive power
flowing periodically into and out of the circuit. In scattering theory,
the real part of this quantity is what is usually referred to as extinction,
and includes all power lost from the incident wave due to scattering
and dissipation processes. The imaginary part corresponding to the
reactance is generally not considered in optical applications, however
given its relationship to the energy stored in near-fields \cite{Vandenbosch2010},
this quantity can yield useful information for metamaterials and nano-photonic
structures, particularly where energy is to be extracted from an emitter.

As shown in the inset of Fig.~\ref{fig:extinction-modes}(a), a plane-wave
incident upon an SRR is considered with its electric field polarized
at $45^{\circ}$ to the gap, with propagation along the ring axis.
In Fig.~\ref{fig:extinction-modes} the contribution of each mode
to the total extinction is plotted, obtained by substituting Eq.~\eqref{eq:modal-inverse}
into Eq.~\eqref{eq:extinction-current}. As this example is for a
lossless structure, the extinction can be attributed entirely to scattering
processes. It can be seen that the line-shapes follow the impedance
given in Fig.~\ref{fig:scalar-admittances}, scaled by the overlap
between the mode and the incident field. The fourth mode is essentially
not excited, and this is consistent with the current distribution
shown in Fig.~\ref{fig:singularities}(b), which has a quadrupolar
type of distribution that does not couple to normally incident plane
waves. It is also clear that the modes which dominate the scattering
process are different from those which dominate the reactive stored
energy, which goes through zero at the complex resonant frequency,
but which can be quite large at other frequencies.

In Fig.~\ref{fig:extinction-comparison} the sum of these modeled
contributions is compared with the direct calculation of the complex
extinction efficiency. The model clearly gives very good agreement
over an extremely wide frequency band, well beyond any quasi-static
circuit limit, or the limit of homogenization if the meta-atom were
placed in a periodic array. This indicates that for this structure
the frequency-dependence of the modal currents can be neglected, whilst
still maintaining good accuracy. It also underlines a potential pitfall
of using dipole moments as the fundamental degrees of freedom, since
modes 1 and 3 have dipole moments parallel to the gap, but are clearly
governed by completely different dynamics.

The computational performance of both the direct calculation and the
model are both dominated by performing the integrations in Eq.~\eqref{eq:impedance_matrix}
to fill the impedance matrix, and for the direct solution solving
it for the incident field. The relative performance of the direct
solution and the model depends on the number of modes, and the number
of frequencies at which the results are calculated. The results shown
in Fig.~\ref{fig:extinction-comparison} were calculated using a
computer with an i7-3740QM 2.7\,GHz quad-core CPU. Searching for
the four singularities and fitting the scalar model takes approximately
40\,s, which enables the extinction cross-section to be calculated
in 0.3\,s. In contrast, the naive approach of directly solving the
system at 500 frequencies takes approximately 444\,s.

\section{Coupling of open resonators\label{sec:coupling-meta-atoms}}

In plasmonic and dielectric oligomers, many interesting effects arise
due to coupling between closely spaced elements. Furthermore, the
electromagnetic response of a metamaterial can differ markedly from
that of an individual meta-atom, due to near-field interaction. The
model of resonant scatterers based on their complex singularities
is an ideal tool to study this coupling.

\subsection{The coupled resonator model}

The hybridization model has been highly successful in describing the
interaction of plasmonic resonators \cite{Prodan2003} and meta-atoms
\cite{Liu2007b,powell_near-field_2011}. In this model hybrid modes
emerge due to quasi-static interaction between the modes of elements.
A Lagrangian of the system is defined, accounting for the stored energy
in the inductance and capacitance of the resonant elements, each of
which is described by the excitation of its fundamental mode. A procedure
for calculating the coefficients of interaction was given in Ref.~\onlinecite{Powell2010},
based on quasi-static calculations of the stored energy. However for
many structures, interaction can be significant even at long distances,
where retardation becomes significant \cite{Decker2011}. Retardation
introduces a phase factor which can greatly change the phase of the
interaction constants \cite{Turner2010,Liu2012a,Tatartschuk2012},
and makes all stored energy quantities complex. This breaks the underlying
physical assumptions of a Lagrangian model, and physical meaning can
only be restored by including the full spectrum of plane waves, as
per the system and bath approach discussed in Section \ref{sec:modes-open-resonators}.

The difficulties arising from the use of the Lagrangian can be avoided
by instead using the EFIE operator of Eq.~\eqref{eq:EFIE}, which
is applicable to an arbitrary number of elements and which naturally
includes all retardation effects. The considerations for modeling
interaction between open resonators are essentially identical to those
for a single resonator discussed in Section \ref{sec:modes-open-resonators}.
The procedure developed in Section \ref{sec:model-single} gives a
compact description of a single element, and needs to be extended
to include the interaction terms. By weighting the mutual parts of
the impedance matrix with the modes of the open resonators, clear
physical meaning can be given to the interactions coefficients, along
with a simple recipe for their calculation. The result is a reduced
matrix equation

\begin{equation}
\sum_{i,m}\left[s\check{L}_{\left\langle i,j\right\rangle }^{\left(m,n\right)}\left(s\right)+\frac{1}{s}\check{S}_{\left\langle i,j\right\rangle }^{\left(m,n\right)}\left(s\right)\right]\check{I}_{\left\langle i\right\rangle }^{\left(m\right)}\left(s\right)=\check{V}_{\left\langle j\right\rangle }^{\left(n\right)}\left(s\right),
\end{equation}
where the angle-bracketed subscript refers to the scatterer, and the
superscript refers to the mode number. The self-terms of this reduced
matrix are taken directly from Eq.~\eqref{eq:scalar-model}, and
the mutual are weighted by the current vectors of the relevant modes

\[
\check{L}_{\left\langle i,j\right\rangle }^{\left(m,n\right)}\left(s\right)=\mathrm{I_{\left\langle i\right\rangle }^{\left(m\right)}}\left(s^{\left(m\right)}\right)\cdot\mathrm{L_{\left\langle i,j\right\rangle }}\left(s\right)\cdot\mathrm{I_{\left\langle j\right\rangle }}^{\left(n\right)}\left(s^{\left(n\right)}\right),
\]
and similarly for $\check{S}_{\left\langle i,j\right\rangle }^{\left(m,n\right)}$.
These coupling terms are similar to those derived in Refs.~\onlinecite{Turner2010,Liu2012a,Tatartschuk2012},
but here they are based on well-defined modes, and an arbitrary number
of modes can be included in the coupling process. The source field
in the reduced model is also obtained by weighting with the mode current
vector

\[
\check{V}_{\left\langle i\right\rangle }^{\left(m\right)}\left(s\right)=\mathrm{I_{\left\langle i\right\rangle }^{\left(m\right)}}\left(s^{\left(m\right)}\right)\cdot\mathrm{V}_{\left\langle i\right\rangle }\left(s\right).
\]
After solving the reduced system, the current solution on each scatterer
will be a superposition of modal terms:

\[
\mathrm{I_{\left\langle i\right\rangle }\left(s\right)}=\sum_{m}\check{I}_{\left\langle i\right\rangle }^{\left(m\right)}\left(s\right)\mathrm{I_{\left\langle i\right\rangle }^{\left(m\right)}}\left(s^{\left(m\right)}\right)
\]

While the inclusion of retardation effects into the coupling coefficients
increases the accuracy of the model, this comes at the cost of making
the coefficients $\check{L}$ and $\check{S}$ frequency dependent.
In Refs.~\onlinecite{Zheng2013a,Liu2012a} these coefficients were
re-calculated for each frequency, which is accurate, but computationally
inefficient. In comparison to modeling the self impedance terms, the
optimal model of the mutual impedance is more dependent on the specific
parameters of the system. For large separation between the elements,
retardation can result in significant oscillation of the coupling
coefficient with frequency, which may be best accounted for via a
multipole expansion of $\mathbf{j}^{(\alpha)}$. On the other hand,
for closely spaced resonators, the effect of retardation can be relatively
weak, so a low-order polynomial can suffice to describe the interaction.

\subsection{Verification}

\begin{figure}
\includegraphics[width=1\columnwidth]{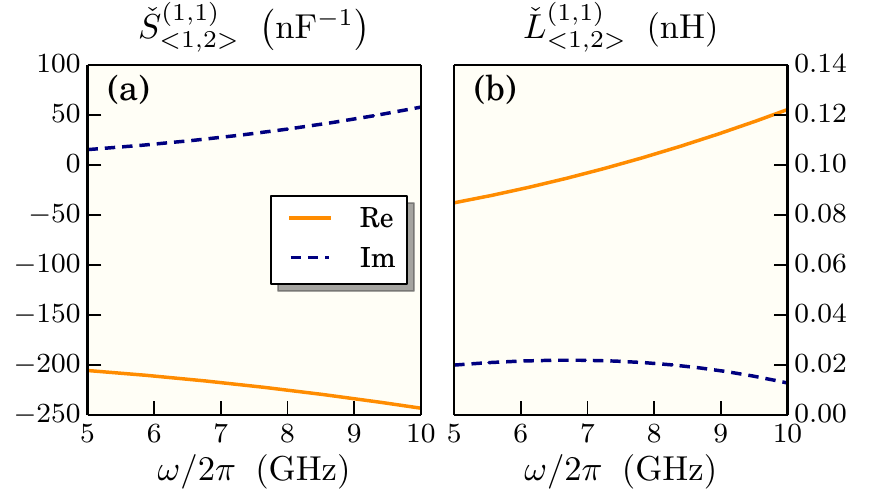}

\protect\caption{(Color online) The (a) capacitive and (b) inductive contributions
to the mutual coupling between a pair of identical SRRs in broadside-coupled
configuration.\label{fig:coupling_terms}}
\end{figure}

The example used to illustrate the coupling problem is a broadside
coupled SRR \cite{marques_role_2002}, consisting of two of the rings
studied in the previous sections, with the second ring rotated by
$180{}^{\circ}$ relative to the first and separated by 2\ mm. This
system is sufficiently simple that its singularities could easily
be found directly, using the same approach as for a single ring. However,
by considering the problem in terms of the modes of individual rings,
the physics of this hybridization process can be shown. In Fig.~\ref{fig:coupling_terms}
the coupling coefficients between the first mode of each ring are
shown. As these functions are very smooth, they are each easily fitted
by a fourth-order polynomial, which requires the mutual part of the
impedance matrix to be filled at 2 different frequencies.

\begin{figure}
\includegraphics[width=1\columnwidth]{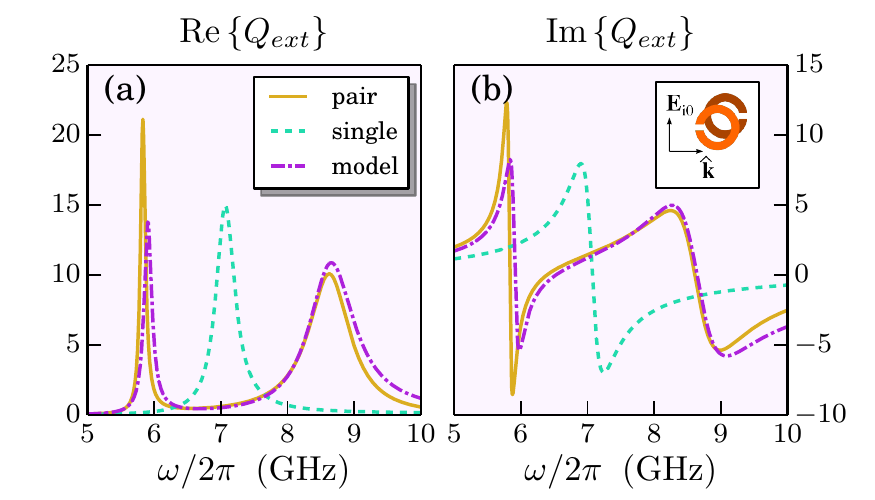}

\protect\caption{(Color online) The (a) real and (b) imaginary parts of the complex
extinction spectrum for a broadside-coupled pair of SRRs, comparing
the model with the directly calculated results. For reference, the
extinction of a single SRR is also shown. The inset gives the polarization
of the incident wave.\label{fig:extinction_bcsrr}}
\end{figure}

It can be seen that with increasing frequency, the imaginary parts
of these coupling coefficients generally becomes more significant,
although the increase is non-monotonic. The differing signs of the
imaginary parts are consistent with the real part of impedance being
positive. The resulting extinction of the broadside coupled SRR pair
is calculated for a plane wave polarized across the gaps, with the
incident magnetic field normal to the rings (recalling that the incident
magnetic field is included in the model implicitly through the gradient
of the electric field). Fig.~\ref{fig:extinction_bcsrr} shows the
corresponding extinction cross-sections, comparing the model with
a direct calculation, and also comparing with a single SRR. As the
polarization and propagation directions of the incident wave are changed,
the fundamental mode of the single ring is excited more strongly than
in Fig.~\ref{fig:extinction-modes}. The splitting of the fundamental
mode is clearly observable. More interestingly, the model shows that
the lower frequency mode, with parallel currents in each ring, has
an enhanced quality factor, corresponding to a reduction in radiative
losses, whereas the higher frequency mode is broadened, due to its
increased radiation efficiency. These effects are directly attributable
to the imaginary parts of $\check{L}_{\left\langle 1,2\right\rangle }^{\left(1,1\right)}$
and $\check{S}_{\left\langle 1,2\right\rangle }^{\left(1,1\right)}$,
shown in Fig.~\ref{fig:coupling_terms}.

\begin{figure}
\includegraphics[width=1\columnwidth]{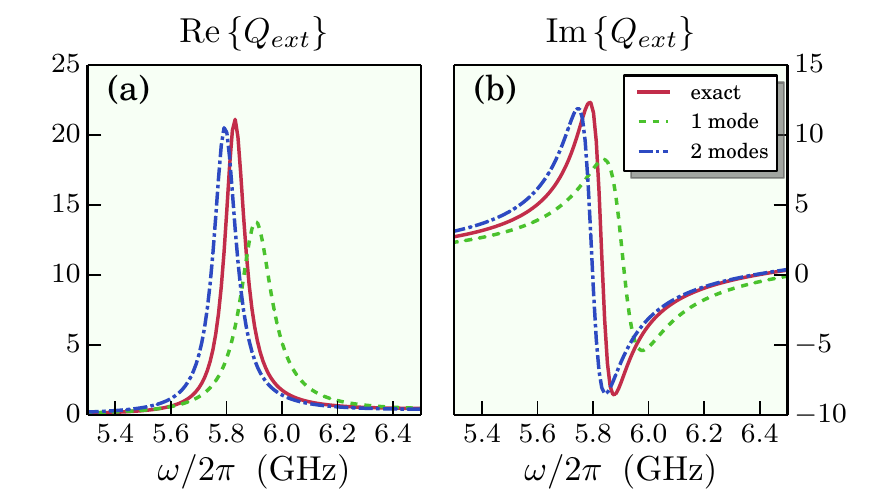}

\protect\caption{(Color online) Improved accuracy of the model of a broadside-coupled
SRR when including an additional mode in the calculation. Comparing
the (a) real part and (b) imaginary part of the models against the
direct calculation.\label{fig:bcsrr_three_modes}}

\end{figure}

It can be seen that the model gives very good agreement with the direct
calculation. This agreement can be improved by considering more than
one mode on each of the rings. The dominant mode of each ring is mode
1 shown in Fig.~\ref{fig:singularities}(b). It is clear from inspection
of mode 2 that it has quite different symmetry to mode 1, and it was
confirmed numerically that it does not play a role in coupling between
rings in this configuration. However, mode 3 also has opposite signs
of the charges across the SRR gap just like mode 1, and it also plays
a role in the formation of the modes of the coupled system. This is
illustrated in Fig.~\ref{fig:bcsrr_three_modes}, which gives a magnified
view of the lower frequency hybridized mode. It can be seen that the
additional mode further increases the accuracy of the model, and clearly
indicates the contribution of this mode to the coupling process. It
was found that the remaining discrepancy is not remedied by increasing
the number of modes in the coupled model. Instead, it is due to the
impedance of the single ring being fitted near its resonance, whereas
the resonances of the coupled system are strongly shifted, to a region
where the fitting is less accurate. Improved accuracy would require
a robust approach to fit a higher-order model than Eq.~\eqref{eq:scalar-model}
to the data, and the frequency dependence of the current eigenvector
$I^{(\alpha)}$ to be accounted for, which have not been achieved
so far. However, the accuracy of the existing approach is likely to
be sufficient for most applications, even if only the dominant mode
is considered.

\section{Conclusion}

In this paper, a technique was presented to describe the physics of
meta-atoms, (nano-)antennas and similar small resonant particles.
Using an integral operator approach allows the radiation boundary
conditions to be modeled efficiently, and the modes of the system
are found by searching for the complex frequencies where this operator
becomes singular. It was shown that this approach remains physically
meaningful and accurate in regimes where dipole and quasi-static models
fail. The coupling of two rings to form a broadside-coupled SRR was
studied, showing that the coupling coefficients are smooth, and can
be calculated efficiently. The calculated interaction constants were
shown to model not only the frequency splitting, but also the enhancement
and suppression of radiation losses for the two modes of the coupled
system. It was demonstrated that the model can readily incorporate
higher-order corrections due to additional modes of each scatterer
contributing to the coupled mode.
\begin{acknowledgments}
I acknowledge fruitful discussions with Yuri Kivshar, Ilya Shadrivov,
Andrey Miroshnichenko, Stanislav Maslovski, Maxim Gorkunov, Constantin
Simovski, Sergei Tretyakov, Pavel Belov, Ben Hopkins, Mingkai Liu,
Jakob de Lasson and Christophe Sauvan. This work was funded by the
Australian Research Council.
\end{acknowledgments}
\appendix

\section{Details of the EFIE\label{sec:EFIE-details}}

The free space Green's function appearing in the electric field integral
equation is given by

\begin{equation}
\overline{\overline{G}}_{0}\left(\mathbf{r}\right)=\left[-s\mu\overline{\overline{I}}+\frac{1}{s\epsilon}\nabla\nabla\right]\frac{\exp(-\gamma|\mathbf{r}|)}{4\pi|\mathbf{r}|}.\label{eq:greens_function}
\end{equation}

The electric field integral equation is tested and weighted to obtain
the impedance matrix $\mathrm{Z}(s)$. After algebraic manipulation
to transfer the gradient operations onto the basis functions, the
elements of the impedance matrix are given by \cite{Gibson2008}

\begin{widetext}

\begin{equation}
Z_{mn}=\iint_{T_{m}}\iint_{T_{n}}\left(s\mu\mathbf{f}_{m}\left(\mathbf{r}\right)\cdot\mathbf{f}_{n}\left(\mathbf{r}'\right)+\frac{1}{s\varepsilon}\left[\nabla\cdot\mathbf{f}_{m}\left(\mathbf{r}\right)\right]\left[\nabla'\cdot\mathbf{f}_{n}\left(\mathbf{r}'\right)\right]\right)\frac{e^{-\gamma\left|\mathbf{r}-\mathbf{r'}\right|}}{4\pi\left|\mathbf{r}-\mathbf{r'}\right|}\mathrm{d^{2}}\mathbf{r}'\mathrm{d}^{2}\mathbf{r},\label{eq:impedance_matrix}
\end{equation}

\end{widetext}where $\varepsilon$ and $\mu$ are the permittivity
and permeability of the background medium.

\section{Implementation details\label{sec:implementation-details}}

The approach proposed in this paper is implemented in an open source
code \cite{openmodes}. Some essential details of the numerical techniques
and computational tools used are given here. The geometry is created
in boundary representation (B-rep) form, and is converted to a triangular
surface mesh using the package \texttt{gmsh} \cite{Geuzaine2009}.
Basis functions are then defined on the mesh, using either rooftops
\cite{Rao1982}, or loops and stars \cite{Vecchi1999}, which give
superior spectral properties for mesh elements which are small compared
to the wavelength. 

The singularities of Eq.~\eqref{eq:impedance_matrix} as $\mathbf{r}\rightarrow\mathbf{r}'$
are integrable, and are accounted for by subtracting the singular
terms from the Green's function and integrating them separately \cite{Arcioni2002,Hanninen2006}.
The remaining non-singular integrals are computed using a fifth order
symmetric integration rule \cite{Dunavant1985}.

The oscillator model given by Eq.~\eqref{eq:scalar-model} is fitted
using the non-negative least squares algorithm, which ensures that
only real coefficients of the correct sign appear in the fitting polynomial
\cite{lawson_solving_1995}. The result is a fitted impedance function
where the real part is positive to satisfy passivity, and the imaginary
part increase with frequency in accordance with Foster's reactance
theorem \cite{foster_reactance_1924}.

The code to implement these methods is written in the Python language,
with the most computationally intensive routines written in Fortran.
It utilizes the scientific python tools \texttt{numpy}, \texttt{scipy},
\texttt{matplotlib} and \texttt{IPython}. \cite{Perez2007,Hunter2007,Oliphant2007,Peterson2009}.

\section{Search procedure for resonances\label{sec:search-procedure}}

Finding the zeros of $\mathrm{Z}(s)$ involves the solution of a transcendental
equation, and it is necessary to use iterative techniques. The iterative
search uses Newton iteration to find the value of $s$ which minimizes
the functional \cite{Ruhe1973}

\[
F(s)=\frac{\mathrm{I}(s)\cdot\mathrm{Z}(s)\cdot\mathrm{I}(s)}{\mathrm{I}(s)\cdot\left[\frac{\mathrm{d}}{\mathrm{d}s}\mathrm{Z}(s)\right]\cdot\mathrm{I}(s)}.
\]
The functional is evaluated numerically, with the derivatives of $\mathrm{Z}$
approximated by the difference between subsequent iterations.

A key requirement for successful application of iterative methods
is a good initial guess, and the approach presented here was empirically
found to be robust. The first step is to decompose the impedance matrix,
as per Eq.~\ref{eq:impedance_LS}. These matrices are evaluated at
some arbitrary initial frequency $s_{i}$, and the linearized problem
is obtained by neglecting frequency variation of these matrices and
solving for the homogeneous solutions $\tilde{s}^{(l)}$ and $\tilde{\mathrm{I}}^{(l)}$
of

\begin{equation}
\mathrm{S}\left(s_{i}\right)\cdot\tilde{\mathrm{I}}^{(l)}=-\left(\tilde{s}^{(l)}\right)^{2}\mathrm{L}\left(s_{i}\right)\cdot\tilde{\mathrm{I}}^{(l)},\label{eq:linearised_eigenvalues}
\end{equation}
which is a generalized eigenvalue problem solvable by standard routines.
However, Eq.~\eqref{eq:linearised_eigenvalues} has non-physical
solution at $s=0$ corresponding to the null space of the scalar potential
part of the impedance operator. These solutions can be eliminated
through the use of loop-star basis functions, which decompose the
current into loops (having zero divergence) and stars (the remaining
component which is almost irrotational) \cite{Vecchi1999}. Each vector
and matrix is partitioned between loop ($\mathrm{l}$) and star ($\mathrm{s}$)
components, where by the zero divergence of the loop basis functions,
Eq.~\eqref{eq:linearised_eigenvalues} becomes:

\begin{equation}
\begin{bmatrix}0 & 0\\
0 & \mathrm{S_{ss}}
\end{bmatrix}\begin{bmatrix}\tilde{\mathrm{I}}_{\mathrm{l}}\\
\tilde{\mathrm{I}}_{\mathrm{s}}
\end{bmatrix}=-\left(\tilde{s}^{(l)}\right)^{2}\begin{bmatrix}\mathrm{L_{ll}} & \mathrm{L_{ls}}\\
\mathrm{L_{sl}} & \mathrm{L_{ss}}
\end{bmatrix}\begin{bmatrix}\tilde{\mathrm{I}}_{\mathrm{l}}\\
\tilde{\mathrm{I}}_{\mathrm{s}}
\end{bmatrix}.
\end{equation}
First note that for $s=0$, any solution with $\tilde{\mathrm{I}}_{\mathrm{s}}=0$
will satisfy this equation. These are the solutions to be eliminated,
whereas for the cases of physical interest $\tilde{\mathrm{I}}_{\mathrm{l}}=-\mathrm{L_{ll}}^{-1}\mathrm{L_{ls}\tilde{\mathrm{I}}_{s}}$.
This leads to the eigenvalue problem

\[
\mathrm{S_{ss}}\tilde{\mathrm{I}}_{\mathrm{s}}=-\left(\tilde{s}^{(l)}\right)^{2}\left[\mathrm{L_{ss}}-\mathrm{L_{ls}\mathrm{L_{ll}}^{-1}L_{ls}}\right]\tilde{\mathrm{I}}_{\mathrm{s}},
\]
which is solved to find the trial solution $\tilde{\mathrm{I}}_{\mathrm{s}}$
and $\tilde{s}^{(l)}$, used to start the Newton iteration procedure\cite{Ruhe1973,Lancaster1966}.

\bibliographystyle{apsrev4-1}
\bibliography{refs}

\end{document}